\documentclass[pra,twocolumn,amsmath,amssymb,floatfix,reprint,footinbib,superscriptaddress,longbibliography,showkeys]{revtex4-2}
\usepackage{dcolumn}
\usepackage{mathtools}
\usepackage{graphicx}
\usepackage{mathrsfs}
\usepackage{mdwlist}
\usepackage{subfigure}
\usepackage{booktabs}
\usepackage{amsmath}
\usepackage{appendix}
\usepackage{extarrows}
\usepackage{dsfont}
\usepackage{amstext}
\usepackage{amssymb}
\usepackage{amsbsy}
\usepackage{bbm}
\usepackage{bm}
\usepackage{amsthm}
\usepackage{graphicx}
\usepackage{textcomp}
\usepackage{multirow}
\usepackage{threeparttable}
\usepackage{soul}
\usepackage{color}
\usepackage{color}
\usepackage{diagbox}
\usepackage[colorlinks,citecolor=blue]{hyperref}
\usepackage{tabularx}
\definecolor{Dgreen}{RGB}{0, 100, 0}
\usepackage{url}
\usepackage[colorlinks]{hyperref}
\hypersetup{%
	plainpages=true,
	breaklinks=true, %not default in dvips mode, so we must specify
	hypertexnames=false, %not ideal, but needed when pagenums duplicate (`i' vs. `1')
	pageanchor=true,
	colorlinks=true,
	linkcolor={blue},
	citecolor={blue},
	urlcolor={blue},
	anchorcolor={black}
}

\begin{document}
	
\title{Enhanced weak-field sensing using detuned motional squeezing}
	
\author{Li-Wen Zhang}
\affiliation{Center for Joint Quantum Studies and Department of Physics, School of Science, Tianjin University, Tianjin 300350, China}
\author{Yang Liu}
\affiliation{Center for Joint Quantum Studies and Department of Physics, School of Science, Tianjin University, Tianjin 300350, China}
\author{Han-Jie Zhu}
\affiliation{Center for Joint Quantum Studies and Department of Physics, School of Science, Tianjin University, Tianjin 300350, China}
\author{Wei Qin}\thanks{Contact author: qin.wei@tju.edu.cn}
\affiliation{Center for Joint Quantum Studies and Department of Physics, School of Science, Tianjin University, Tianjin 300350, China}
\affiliation{Tianjin Key Laboratory of Low Dimensional Materials Physics and Preparing Technology, Tianjin University,
	Tianjin 300350, China}
	
\begin{abstract}
Achieving measurement precision beyond the standard quantum limit (SQL) is a fundamental object in quantum metrology. In this manuscript, we propose a streamlined protocol for weak signal sensing in a trapped-ion spin-oscillator system that simultaneously exploits detuned motional squeezing and spin-oscillator entanglement. By combining these two quantum resources, the phase accumulation in this spin system is significantly accelerated, as a result achieving an exponential improvement in the sensitivity over unsqueezed cases. Remarkably, we show that the significant metrological gain provided by our protocol persists in noisy environments, with the sensitivity remaining below the SQL. In particular, our protocol retains a distinct advantage in the short interrogation time regime, thus demonstrating the practical potential in quantum-enhanced sensing. 
\end{abstract}
	
\maketitle
	
\section{Introduction}

Quantum metrology exploits fundamental principles of quantum mechanics to enable high precision measurements~\cite{PhysRevLett.96.010401,doi:10.1142/S0219749909004839,RevModPhys.89.035002,doi:10.1126/science.aaw2884,doi:10.1126/sciadv.adw9757,Giovannetti2011,Len2022,Greve2022}. In contrast to traditional measurements, it can surpass the standard quantum limit (SQL)~\cite{doi:10.1126/science.1104149}, and improve the sensitivity even to approach the Heisenberg limit (HL) by using quantum resources. These capabilities have led to a wide range of applications in frontier areas of physics, including gravitational wave detection~\cite{PhysRevLett.132.190001,PhysRevLett.121.031101,PhysRevLett.110.171102,PhysRevX.13.021019} and atomic clocks~\cite{cy84-v69x,Nichol2022,Pedrozo-Peñafiel2020}. 

Quantum parameter estimation, a central task in quantum metrology, integrates estimation theory with quantum mechanics to enable the precise inference of physical parameters, such as phase, frequency, and displacement~\cite{Huang2024,Colombo2022,PhysRevLett.123.073001,LI2020126413}. Quantum resources like quantum entanglement and squeezing~\cite{Scully_Zubairy_1997,PhysRevLett.129.123602,PhysRevLett.133.233605,PRXQuantum.5.020314,McCormick2019} can significantly enhance parameter estimation precision, ultimately enabling sensitivities beyond the SQL. For displacement estimation, the SQL is $1/2$, whereas for phase estimation it scales as $1/2\sqrt{\bar{n}}$, where $\bar{n}$ denotes the number of particles involved in the measurement~\cite{PhysRevA.73.023803,PhysRevA.54.R4649,Wang2019,PhysRevA.94.022313,Deng2024,PhysRevLett.105.180402}. So far, the sub-SQL parameter estimation has been experimentally demonstrated using either quantum entanglement or squeezing~\cite{PhysRevLett.132.220801,PhysRevA.107.032425,doi:10.1126/science.abi5226}.  More recently, proposals based on their sequential~\cite{PhysRevA.107.032425} and simultaneous~\cite{PhysRevLett.132.220801} use have been proposed. In particular, the latter can yield a higher sensitivity for electronic field sensing, i.e., displacement estimation, compared to the case of using quantum entanglement or squeezing alone. However, this proposal needs multiple squeezing and anti-squeezing operations, which increases operational complexity and inevitably leads to additional errors in practical implementations~\cite{PhysRevLett.132.220801,PhysRevA.107.032425,doi:10.1126/science.abi5226}.

To overcome this limitation, in this manuscript we develop a simple protocol for displacement estimation, by simultaneously exploiting detuned motional squeezing and spin-oscillator entanglement in a trapped-ion system. Trapped-ion platforms have been widely employed for parameter estimation in hybrid quantum interferometers, owing to their high sensitivity to weak signals and excellent controllability of both internal spin states and motional modes~\cite{doi:10.1126/science.abi5226,doi:10.1126/science.aaf3397,PhysRevLett.121.040503,PhysRevA.102.052609,Wolf2019}.

In our proposal, the signal of a weak field is mapped onto a spin phase shift, mediated by a Bogoliubov mode of the motional degree of freedom, rather than the bare mode as usual. The accumulated phase is then extracted through interferometric spin-state readout~\cite{Maiwald2009,PhysRevLett.105.013004,Biercuk2010}. By appropriately engineering the system, we find that squeezing simultaneously enhances both the spin-motion coupling and the weak field signal to be measured. At the same time, a tunable detuning of the motional resonance from the signal field resonance optimizes the interrogation time. Compared to the previous scheme, our method can enable a higher measurement sensitivity, with fewer operations and a shorter interrogation time. Specifically, as seen below, the scheme in Ref.~\cite{PhysRevLett.132.220801} requires {\it eight} successive operations, but our approach only needs a {\it single} operation, thus reducing the potential sources of error and substantially improving the experimental feasibility. Moreover, for modest parameters, our proposal also yields an enhancement of $\sim25\%$ in the sensitivity during a short timescale.

The remaining of the paper is organized as follows. The physical system used for measurement, i.e., a single trapped-ion system, is introduced in Sec.~\ref{physical model}, including the corresponding system Hamiltonian and the state evolution. Sec.~\ref{enhanced sensitivity} introduces the squeezing enhanced sensitivity measurement in our scheme. Sec.~\ref{noise-affected system} analyzes the noise effect and shows that the SQL can still be surpassed, yielding a significant gain. Conclusions are drawn in Sec.~\ref{Conclusion}, where we show that our scheme has a clear advantage in the short-time regime. Finally, the Appendix provide some detailed derivations.
\section{physical model}
\label{physical model}
\begin{figure}[t] 
	\centering
	\includegraphics[width=9cm]{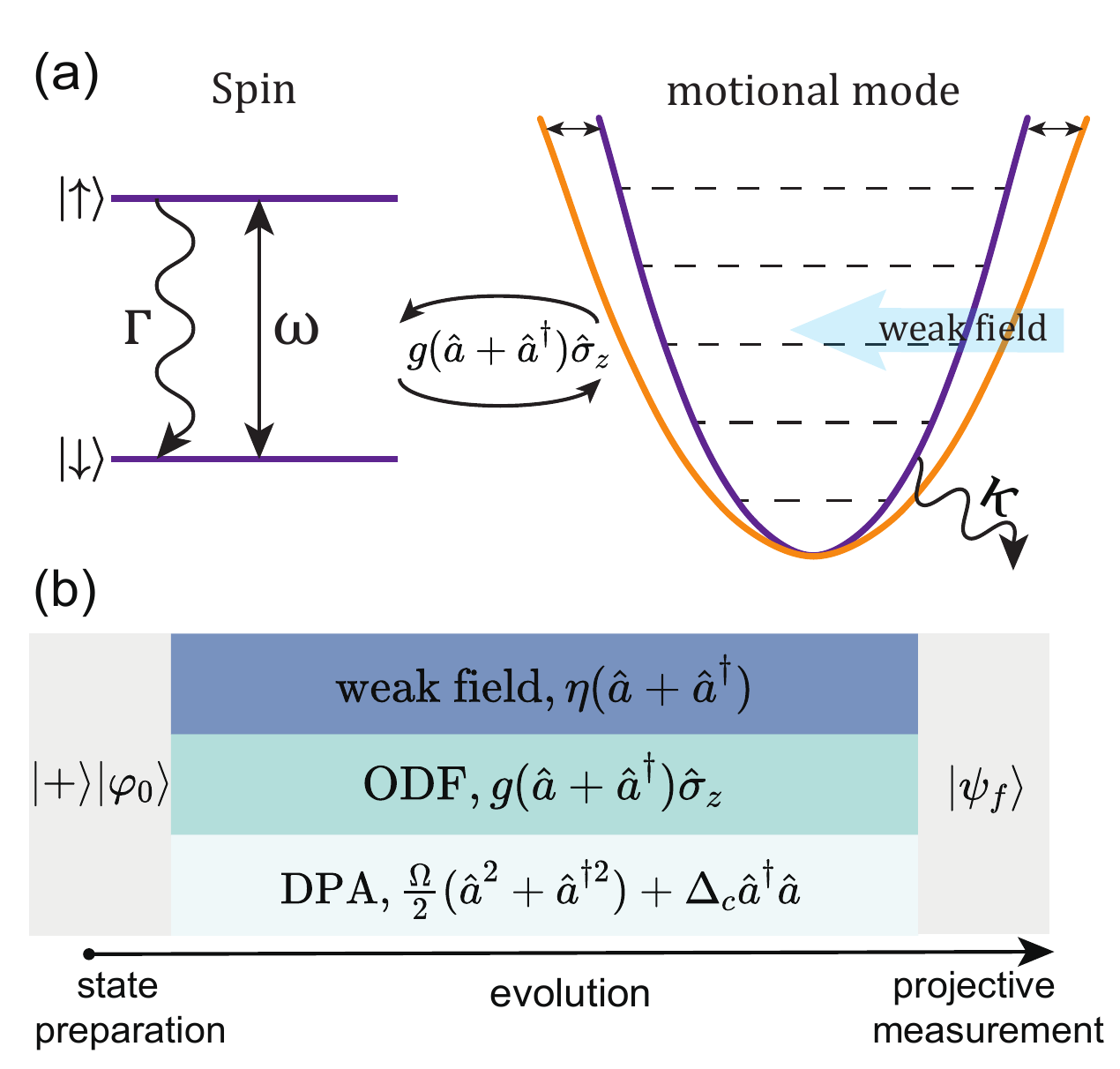}
	\caption{(a) Schematic of the trapped-ion system used in our protocol. The spin and the motional mode are entangled through a spin-dependent ODF with strength $g$, while squeezing the motional mode is simultaneously applied to enhance both the coupling strength and the weak field to be measured. The spin has two states $\vert{\uparrow}\rangle$ and $\vert{\downarrow}\rangle$, with a transition frequency $\omega$ and a decay rate $\Gamma$. The damping rate of the motional mode is denoted by $\kappa$. (b) Our single-step protocol for enhancing the sensitivity of a weak field, based on detuned motional squeezing and spin-motion entanglement.}
	\label{tensor}
\end{figure}
The basic idea underlying our proposal is schematically depicted in Fig.~\ref{tensor}. Specifically, we consider, as in Fig.~\ref{tensor}(a), a trapped-ion system consisting of a harmonic oscillator, as a motional mode of frequency $\omega_{c}$, coupled to a two-level spin through a spin-dependent optical dipole force (ODF) of strength $g$ and phase $\phi_{D}$~\cite{PhysRevA.100.043417}. We further assume that this motional mode is subject to a detuned parametric amplification (DPA). The system can thus be described, in a proper rotating frame, by the following Hamiltonian (hereafter, $\hbar=1$),
\begin{align}\label{full_Hamil}
	\hat{H}=\;&\Delta_c\hat{a}^\dagger\hat{a}+g(e^{-i\phi_D}\hat{a}+e^{i\phi_D}\hat{a}^{\dagger})\hat{\sigma}_z\nonumber\\
	\;&+\frac{\Omega}{2}(e^{i\theta}\hat{a}^{\dagger2}+e^{-i\theta}\hat{a}^2)+\eta(e^{-i\phi_E}\hat{a}+e^{i\phi_E}\hat{a}^\dagger),
\end{align}
where $\hat{\sigma}_z$ is the Pauli matrix of the spin, $\hat{a}$ $(\hat{a}^\dagger)$ is the annihilation (creation) operator of the motional mode, $\Omega$ and $\theta$ are the amplitude and phase of the DPA, respectively. Moreover, $\eta$ and $\phi_E$ refer to the strength and phase of the signal field, respectively. This signal field is considered as an external force acting on the motional mode, e.g., an oscillating electric field~\cite{Wolf2019}, and $\Delta_c=\omega_{c}-\omega_{d}$ is a detuning of the motion from the signal field resonance $\omega_{d}$. Figure~\ref{tensor}(b) illustrates our weak-field sensing protocol. We consider an initial state $\vert+\rangle\vert\varphi_0\rangle$, where $\vert+\rangle=\frac{1}{\sqrt{2}}(\vert{\uparrow}\rangle +\vert{\downarrow}\rangle)$ is an equal superposition of the two spin states $\vert{\uparrow}\rangle$ and $\vert{\downarrow}\rangle$, and $\vert\varphi_0\rangle$ represents an arbitrary motional state. Our proposal simultaneously squeezes the motional mode under detuning, entangles the spin and motional degrees of freedom, and accumulates the spin phase induced by the weak signal to be measured. In this case, the weak field is mapped onto the system dynamics during the evolution and ultimately onto the spin phase.

For simplicity, we consider the case $\phi_D=\phi_E=0$, and the general case is discussed in Appendix~\ref{Derivation of the state}. We then introduce a Bogoliubov mode $\hat{\beta}=\cosh(r)\hat{a}+e^{i\theta}\sinh(r)\hat{a}^\dagger$, with $r=\frac{1}{4}\ln[(\Delta_c+\Omega)/(\Delta_c-\Omega)]$ being the squeezing degree~\cite{Bogoljubov1958,PhysRevLett.120.093601,3mlc-r8x4}. Consequently, we have $\Delta_c\hat{a}^\dagger\hat{a}+\frac{\Omega}{2}(e^{i\theta}\hat{a}^{\dagger2}+e^{-i\theta}\hat{a}^2)=\omega_{\text{sq}}\hat{\beta}^\dagger\hat{\beta}$, and the Hamiltonian $\hat{H}$ is likewise transformed to
\begin{align}\label{Bgl Hamil no reduct}
	\hat{H}_{\text{sq}}=\;&\omega_{\text{sq}}\hat{\beta}^\dagger\hat{\beta}+g(\alpha^*\hat{\beta}+\alpha\hat{\beta}^{\dagger})\hat{\sigma}_z+\eta(\alpha^*\hat{\beta}+\alpha\hat{\beta}^{\dagger}).
\end{align}
Here, $\omega_{\text{sq}}=\sqrt{\Delta_c^2-\Omega^2}$ is the resonance frequency of the mode $\hat{\beta}$, and $\alpha=\cosh(r)-e^{i\theta}\sinh(r)$ is an amplification factor. It can be seen from Eq.~\eqref{Bgl Hamil no reduct} that setting $\theta=\pi$ gives a maximum amplification of the ODF and the electric field. In this case, the Hamiltonian takes the form
\begin{align}\label{Bogoliubov Hamiltonian}
	\hat{H}_{\text{sq}}=\;&\omega_{\text{sq}}\hat{\beta}^\dagger\hat{\beta}+ge^r(\hat{\beta}+\hat{\beta}^{\dagger})\hat{\sigma}_z+\eta e^r(\hat{\beta}+\hat{\beta}^{\dagger}).
\end{align}
The state $|\psi(t)\rangle$ at an arbitrary evolution time $t$ can then be written as (see Appendix~\ref{Derivation of the state})
\begin{align}\label{state_with_t}
	\vert\psi(t)\rangle
	=e^{i\phi(t)\hat{\sigma}_z}\hat{D}_\beta[\alpha(t)]e^{-i\omega_{\text{sq}}\hat{\beta}^\dagger\hat{\beta} t}\vert+\rangle\vert\varphi_0\rangle.
\end{align}
Here,
\begin{equation}\label{Dbeta}
	\hat D_\beta[\hat{\alpha}(t)]
	=
	\exp\left[
	\hat{\alpha}(t)\hat\beta^\dagger
	-
	\hat{\alpha}^\dagger(t)\hat\beta
	\right]
\end{equation}
is the displacement operator of the Bogoliubov mode, and
\begin{align}
	\hat{\alpha}(t)&=\frac{-e^r(\eta+g\hat{\sigma}_z)(1-e^{-i\omega_{\text{sq}}t})}{\omega_{\text{sq}}},\label{alpha(t)}\\
	\phi(t)&=\frac{2\eta ge^{2r}(\omega_{\text{sq}}t-\sin(\omega_{\text{sq}}t))}{\omega_{\text{sq}}^2}\label{phi(t)}.
\end{align}
We further choose the frequency of the Bogoliubov mode to be 
\begin{align}
\omega_{\text{sq}}=2\pi/T,
\end{align}
where $T$ is the interrogation time, so that the spin-oscillator system becomes disentangled at $t=T$~\cite{Leibfried2003,Burd2021}. Consequently, the final state in Eq.~\eqref{state_with_t} is reduced to
\begin{align}\label{reduced final state}
	\vert\psi_f\rangle&=\exp[i\phi(T)\hat{\sigma}_z]\vert+\rangle\vert\varphi_0\rangle,
\end{align}
with 
\begin{equation}\label{e2r}
	\phi(T)
	=
	\phi_0(T)e^{2r}.
\end{equation}
Here, $\phi_{0}(T)=\eta gT^2/\pi$ is the accumulated phase with no squeezing for the same interrogation time $T$. It is seen that by mapping the motional displacement caused by the weak signal $\eta$ onto the spin degree of freedom, the signal is encoded into a spin phase shift. This accumulated phase is then read out through measuring the $y$ component of the spin, given by
\begin{align}\label{sigmay_f} \langle\hat{\sigma}_y\rangle_f=\langle\psi_f\vert\hat{\sigma}_y\vert\psi_f\rangle=-\sin[2\phi(T)].
\end{align}
It is clear that the amount of the accumulated phase is amplified exponentially with twice the squeezing degree $r$. The exponential enhancement as in Eq.~\eqref{e2r} remains unchanged for an arbitrary phase $\phi_{E}$, provided that the phases $\phi_{D}$ and $\theta$ are adjusted accordingly (see Appendix~\ref{Derivation of the state}).

Figure~\ref{process}(a) shows the Wigner functions of the bare motional mode $\hat a$ at different evolution times.  We take the initial motional state to be the squeezed vacuum state,
\begin{equation}\label{ini_motional_state}
	|\varphi_0\rangle
	=
	\hat{S}(re^{i\theta})|0\rangle
	=
	|0\rangle_\beta,
\end{equation}
where $\hat S(re^{i\theta})$ is the squeezing operator with squeezing degree $r$ and phase $\theta$, and $|0\rangle_\beta$ denotes the vacuum of the Bogoliubov mode. The reduced density matrix used to calculate these Wigner functions is given by
\begin{equation}
		\hat\rho_a(t)
		=
		\operatorname{Tr}_{\rm s}
		\left[
		|\psi(t)\rangle\langle\psi(t)|
		\right].
\end{equation}
where $\operatorname{Tr}_{\rm s}$ denotes the partial trace over the spin degree of freedom. The Wigner function is then evaluated as
\begin{equation}
	W(x,p,t)
	=
	\frac{1}{2\pi}
	\int_{-\infty}^{+\infty} dy\,
	\left\langle x-\frac{y}{2}\right|
	\hat\rho_a(t)
	\left|x+\frac{y}{2}\right\rangle
	e^{ipy}.
\end{equation}

In the Bogoliubov-mode phase space, the motional mode follows two different circular trajectories with radii $e^r(g+\eta)/\omega_{\rm sq}$ and $e^r(g-\eta)/\omega_{\rm sq}$, respectively, when the spin is in the state $\vert{\uparrow}\rangle$ and the state $\vert{\downarrow}\rangle$~\cite{QIN20241}. Expressed in terms of the bare motional mode, the displacement operator in Eq.~\eqref{Dbeta} becomes
\begin{align}
	\hat D_\beta[\hat\alpha(t)]
	=
	\exp\Bigg\{
	&\frac{e^r}{2}
	\left[
	\hat\alpha(t)+\hat\alpha^\dagger(t)
	\right]
	(\hat a^\dagger-\hat a)
	\nonumber\\
	&+
	\frac{e^{-r}}{2}
	\left[
	\hat\alpha(t)-\hat\alpha^\dagger(t)
	\right]
	(\hat a^\dagger+\hat a)
	\Bigg\}.
\end{align}
It is seen that the first term provides an additional factor of $e^r$. Since $\hat\alpha(t)$ already contains a factor of $e^r$ as given in Eq.~\eqref{alpha(t)}, this term therefore carries an overall factor of $e^{2r}$. At $t=T/2$, the motional mode in the bare-mode phase space reaches the maximum displacements,
\begin{align}
	d_+&=\frac{2(g+\eta)e^{2r}}{\omega_{\rm sq}},\\
	d_-&=\frac{2(g-\eta)e^{2r}}{\omega_{\rm sq}},
\end{align}
corresponding to the spin in the state $|{\uparrow}\rangle$ and the state $|{\downarrow}\rangle$, respectively.

The corresponding trajectories in the bare-mode phase space are elliptical, and their semimajor and semiminor axes are
\begin{align}
	R_{{\rm maj},\pm}
	&=
	\frac{d_\pm}{2}
	=
	\frac{(g\pm\eta)e^{2r}}{\omega_{\rm sq}},\\
	R_{{\rm min},\pm}
	&=
	\frac{g\pm\eta}{\omega_{\rm sq}},
\end{align}
respectively. At $t=T$, each trajectory forms a full loop in phase space, and the enclosed areas are
\begin{align}
  	S_+&=\pi R_{{\rm maj},+}R_{{\rm min},+}=\frac{\pi(g+\eta)^2e^{2r}}{\omega_{\rm sq}^2},\\
  	S_-&=\pi R_{{\rm maj},-}R_{{\rm min},-}=\frac{\pi(g-\eta)^2e^{2r}}{\omega_{\rm sq}^2}.
\end{align}

The resulting area difference is the accumulated spin phase, i.e., $S_+-S_-=\phi(T)$, which, as seen in Fig.~\ref{process}(b), induces a rotation of the spin state about the $z$ axis. This spin phase shift can then be read out through a spin-state measurement. Note that, here, in order to plot Fig.~\ref{process}(a), we have chosen the initial motional state to be a squeezed vacuum state. 

\begin{figure}[t] 
	\centering
	\includegraphics[width=9cm]{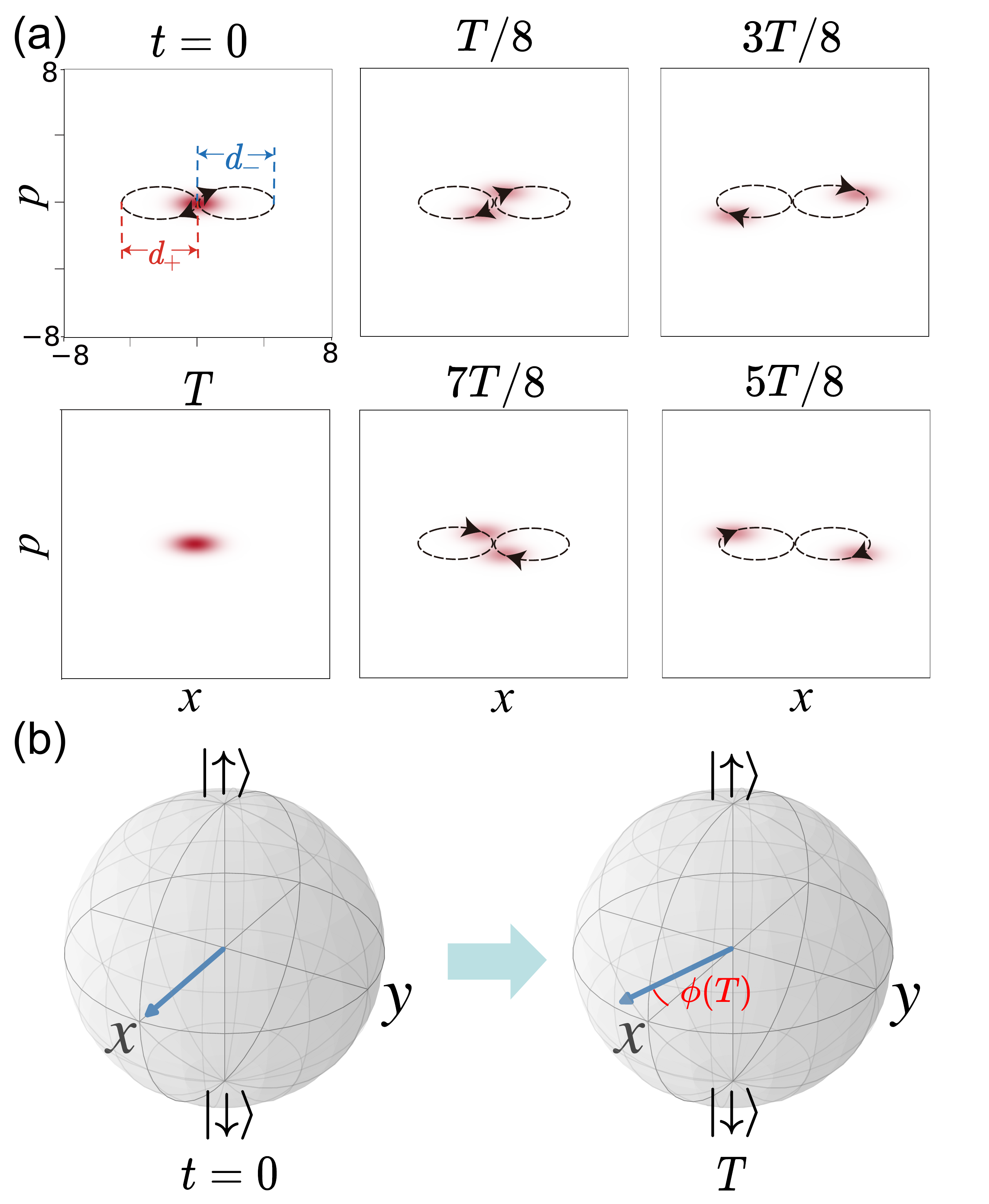}
	\caption{(a) Wigner functions of the motional mode during one complete phase-space loop. The initial state is a squeezed vacuum state $\hat{S}(re^{i\theta})\vert0\rangle$. At $t=T$, the motional state returns to its initial phase-space position, and at the same time the spin and motional degrees of freedom become disentangled. The parameters are $\Omega/2\pi=5~{\rm kHz}$, $g/2\pi=4~{\rm kHz}$, $\eta=100~{\rm Hz}$, and $T=0.2~{\rm ms}$. (b) Schematic representation of the spin states at $t=0$ (left) and $t=T$ (right) on the Bloch spheres. The spin-motion entanglement maps the displacement induced by the signal into a rotation of the spin state, $\hat{U}=e^{i\phi(T)\hat{\sigma}_z}$.}
	\label{process}
\end{figure}

\section{enhanced sensitivity}
\label{enhanced sensitivity}

To quantify the sensitivity enhancement of our protocol, we employ the quantum Fisher information (QFI), $\mathcal{F}^\epsilon_Q(t)$, where $\epsilon$ denotes the parameter to be estimated~\cite{Huang2024}. The ultimate achievable precision is then determined by the QFI through the quantum Cramér-Rao bound (QCRB)~\cite{FUJIWARA1995119,PhysRevLett.72.3439,Gietka2021adiabaticcritical,Helstrom1969,PhysRevLett.124.193602}.
\begin{align}\label{QCRB}
	\Delta \epsilon \geq\frac{1}{\sqrt{\nu\mathcal{F}^\epsilon_Q(t)}},
\end{align}
where $\nu$ represents the number of repeated measurements. For a pure state $\vert\varsigma(t)\rangle$, the QFI can be expressed as
\begin{align}\label{QFI}
	\mathcal{F}^\epsilon_Q(t)=4[\langle\partial_\epsilon\varsigma(t)\vert\partial_\epsilon\varsigma(t)\rangle-\big|\langle\partial_\epsilon\varsigma(t)|\varsigma(t)\rangle\big|^2].
\end{align}
In our protocol, the parameter to be estimated is the weak-field amplitude $\eta$. For the initial motional state $|\varphi_0\rangle$ in Eq.~\eqref{ini_motional_state}, evaluating the QFI with respect to \(\eta\) for the time-dependent state in Eq.~\eqref{state_with_t} yields (see Appendix~\ref{Derivation of the quantum Fisher information})
\begin{align}\label{time-dependent QFI}
\mathcal{F}^\eta_Q(t)
={}&
\frac{16e^{2r}}{\omega_{\mathrm{sq}}^2}
\sin^2\left(\frac{\omega_{\mathrm{sq}}t}{2}\right)
\nonumber\\
&+
\frac{16g^2e^{4r}}{\omega_{\mathrm{sq}}^4}
\left[
\omega_{\mathrm{sq}}t
-\sin(\omega_{\mathrm{sq}}t)
\right]^2.
\end{align}

\begin{figure}[t]
	\centering
	\includegraphics[width=9.0cm]{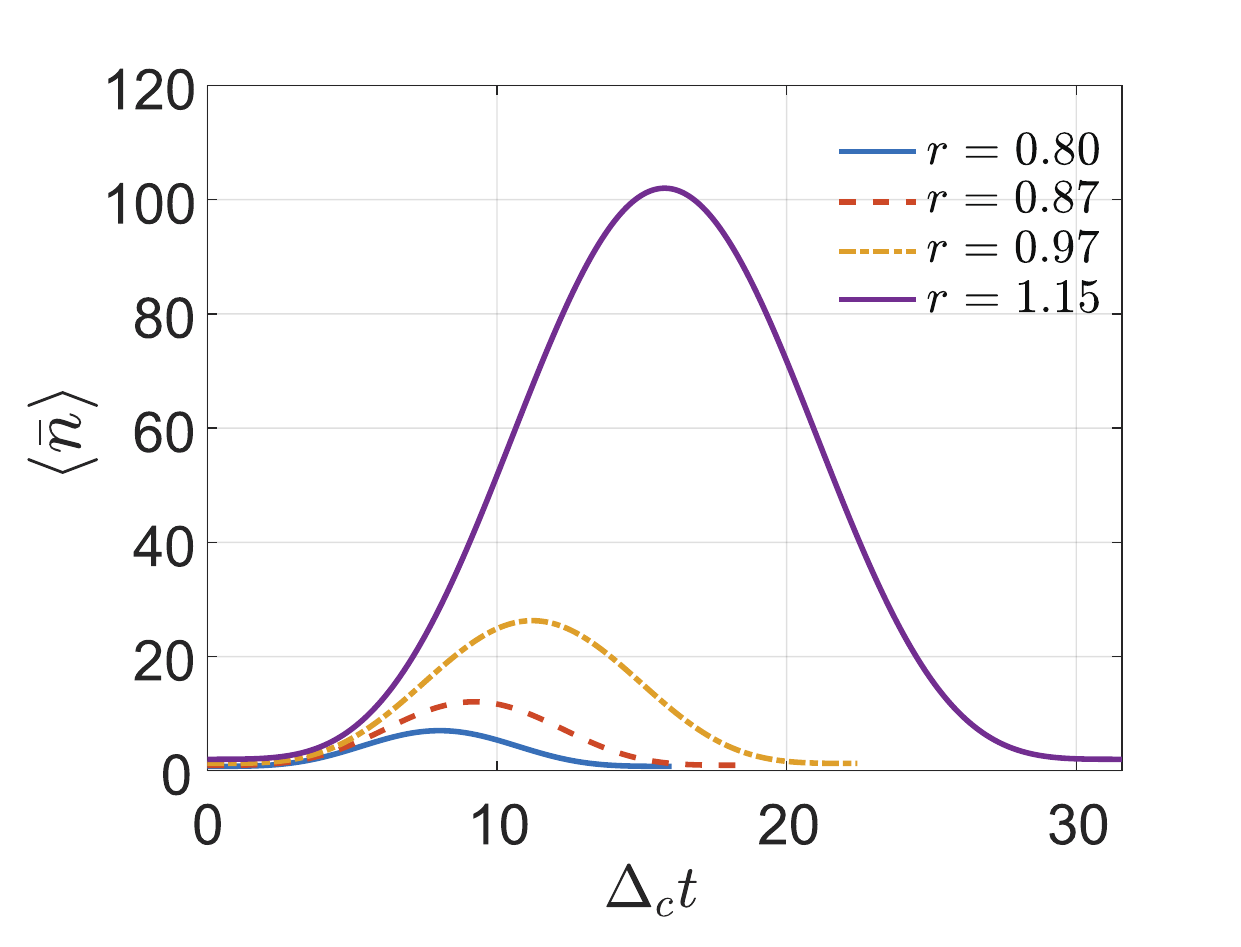}
	\caption{Time evolution of the mean phonon number $\langle\hat{a}^\dagger\hat{a}\rangle$ for a fixed detuning $\Delta_c$ and different squeezing degrees $r$ over a single phase-space loop. The mean phonon number always reaches its maximum value $\bar{n}_{\text{max}}$ at $t=T/2$. We have assumed that $g=0.1\Delta_c$ and $\Delta_c/2\pi=1.0$~kHz.}
	\label{n_for_diff_r}
\end{figure}

To compare the QFI of our protocol with the fundamental limit for displacement sensing, we characterize the maximum mean phonon number available during the evolution. To exclude the contribution from the unknown signal, we set $\eta=0$ in this calculation. The mean phonon number reaches its maximum at $t=T/2$, where the spin-motion state is~\cite{PhysRevLett.132.220801}
\begin{align}\label{Psi}
	\vert\Psi\rangle=\frac{1}{\sqrt{2}}[\vert{\uparrow}\rangle\hat{D}(-x)\hat{S}(re^{i\theta})\vert0\rangle+\vert{\downarrow}\rangle\hat{D}(x)\hat{S}(re^{i\theta})\vert0\rangle],
\end{align}
with $x=gTe^{2r}/\pi$. The corresponding maximum mean phonon number is
\begin{align}\label{phonon number}
	\bar{n}_{\mathrm{max}}
	&=\langle\Psi\vert\hat{a}^\dagger\hat{a}\vert\Psi\rangle
	\nonumber\\
	&=\left(\frac{gTe^{2r}}{\pi}\right)^2+\sinh^2(r).
\end{align}
Figure~\ref{n_for_diff_r} shows the corresponding time evolution of the mean phonon number for different squeezing degrees. Reference~\cite{PhysRevLett.132.220801} evaluates the QFI using a state that does not exist during the corresponding sensing process. To avoid this possible ambiguity, we calculate the QFI directly from the time-dependent state in Eq.~\eqref{state_with_t}, where the weak field is encoded throughout the evolution. Defining the signal-induced displacement as $\zeta=\eta T$, the corresponding QFI is
\begin{equation}\label{QFI_zeta}
	\mathcal{F}^\zeta_Q(t)
	=
	\frac{\mathcal F^\eta_Q(t)}{T^2}.
\end{equation}
As shown in Appendix~\ref{Derivation of the quantum Fisher information}, the QFI of displacement sensing is bounded by~\cite{mhvv-pr5d,PhysRevLett.132.220801}.
\begin{equation}\label{QFI_resource_bound}
	\mathcal{F}^\zeta_Q(t)
	\leq
	16\bar n_{\max}+8,
	\qquad
	0\leq t\leq T.
\end{equation}
Combining Eq.~\eqref{QFI_resource_bound} with the QCRB gives the following lower bound on the displacement sensitivity:
\begin{equation}\label{HL_zeta}
	\Delta\zeta
	\geq
	\frac{1}{4\sqrt{\bar n_{\max}+1/2}}.
\end{equation}
We refer to the right-hand side of Eq.~\eqref{HL_zeta} as the Heisenberg limit for displacement sensing. Compared with the SQL, $(\Delta\zeta)_{\rm SQL}=1/2$, this bound exhibits a $\sqrt{\bar n_{\max}}$ enhancement in the large-resource regime~\cite{mhvv-pr5d,Deng2024}.

In realistic experiments, however, the theoretical precision bound in Eq.~\eqref{HL_zeta} is generally difficult to attain directly. A suitable observable is therefore required to approach this bound~\cite{Huang2024}. In our protocol, we choose the Pauli operator $\hat{\sigma}_y$ for readout. Experimentally, the expectation value of this observable can be obtained via a projective measurement of the spin state~\cite{doi:10.1126/science.abi5226}. The corresponding sensitivity is then quantified using the error-propagation formula from quantum estimation theory~\cite{PhysRevLett.121.043604,PhysRevA.106.022413,RevModPhys.90.035006},
\begin{align}\label{sens_nodiss_zeta}
	\Delta\zeta=\;\frac{\Delta\langle\hat{\sigma}_y\rangle_f}{|\partial\langle\hat{\sigma}_y\rangle_f/\partial\zeta|}=\frac{1}{gT}\frac{\pi}{2e^{2r}}.
\end{align}
Here, $\Delta\langle\hat{\sigma}_y\rangle_f=\sqrt{\langle\hat{\sigma}_y^2\rangle_f-\langle\hat{\sigma}_y\rangle_f^2}$ denotes the standard deviation of $\hat{\sigma}_y$ evaluated for the final state, and $\langle\hat{\sigma}_y\rangle_f$ is given by Eq.~\eqref{sigmay_f}. Using the expression for $\bar{n}_{\text{max}}$ given in Eq.~\eqref{phonon number}, Eq.~\eqref{sens_nodiss_zeta} can be expressed as
\begin{equation}
	\Delta\zeta
	=
	\frac{1}{
		2\sqrt{\bar n_{\mathrm{max}}-\sinh^2r}
	}.
\end{equation}
When the displacement-induced phonon number is much larger than the
squeezing-induced phonon number, i.e.,
\begin{equation}
	\left(\frac{gTe^{2r}}{\pi}\right)^2
	\gg
	\sinh^2r,
\end{equation}
the sensitivity can be approximated as
\begin{align}\label{sens_nodiss_zeta_for n}
	\Delta\zeta\approx\frac{1}{2\sqrt{\bar{n}_{\text{max}}}}.
\end{align}
It is seen that, although the achievable sensitivity does not reach the ultimate theoretical precision bound given in Eq.~\eqref{HL_zeta}, it exhibits the same scaling as $1/\sqrt{\bar{n}_{\text{max}}}$. Moreover, from Eq.~\eqref{sens_nodiss_zeta}, our protocol yields an amplification factor of $2e^{2r}/\pi$, compared to the unsqueezed case in Ref.~\cite{doi:10.1126/science.abi5226}.  

\begin{figure}[t]
	\centering
	\includegraphics[width=9.0cm]{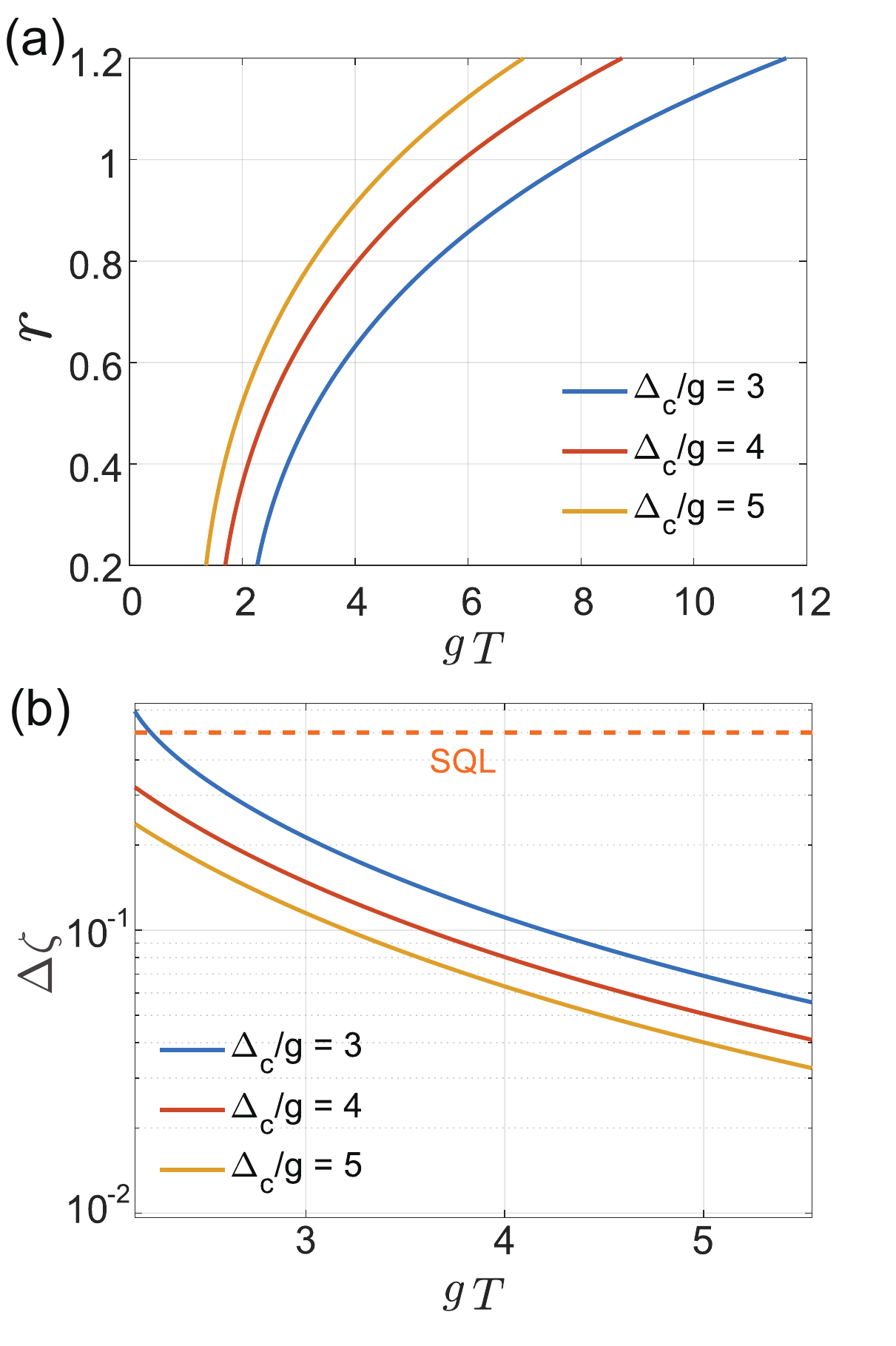}
	\caption{(a) Squeezing degree $r$ and (b) displacement sensitivity $\Delta\zeta$ as a function of the scaled interrogation time $gT$ for different detunings $\Delta_c$. The dashed line in (b) refers to the SQL. The squeezing degree and the sensitivity depend on the ratio of $\Delta_c/g$.}
	\label{diff_deltac}
\end{figure}

To gain further insight into this sensitivity, we now examine how it depends on the squeezing degree $r$ and the interrogation time $T$. For a given detuning $\Delta_c$, the DPA strength can be expressed as $\Omega=\lambda\Delta_c$, with $0\textless\lambda\textless1$. Since
\begin{equation}
		r=\frac{1}{4}\ln\left(\frac{\Delta_c+\Omega}{\Delta_c-\Omega}\right),
		\qquad
		T=\frac{2\pi}{\sqrt{\Delta_c^2-\Omega^2}},
\end{equation}
we obtain the squeezing degree
\begin{align}\label{r}
	r=\frac{1}{4}\ln(\frac{1+\lambda}{1-\lambda}),
\end{align}
and the interrogation time
\begin{align}\label{T}
	T=\frac{2\pi}{\Delta_c\sqrt{1-\lambda^2}}.
\end{align}
Clearly, the squeezing degree $r$ is determined by $\lambda$, while the interrogation time $T$ depends on both $\lambda$ and $\Delta_c$. Therefore, for a given $T$, achieving a larger $r$ requires a larger detuning $\Delta_c$, as shown in Fig.~\ref{diff_deltac}(a). It is further found from Fig.~\ref{diff_deltac}(b) that for a given $T$, a larger $\Delta_c$ yields a larger $r$ and then a higher sensitivity, as a result surpassing the SQL readily.

\section{Effects of technical noise}
\label{noise-affected system}
\begin{figure}[t]
	\centering
	\includegraphics[width=9.0cm]{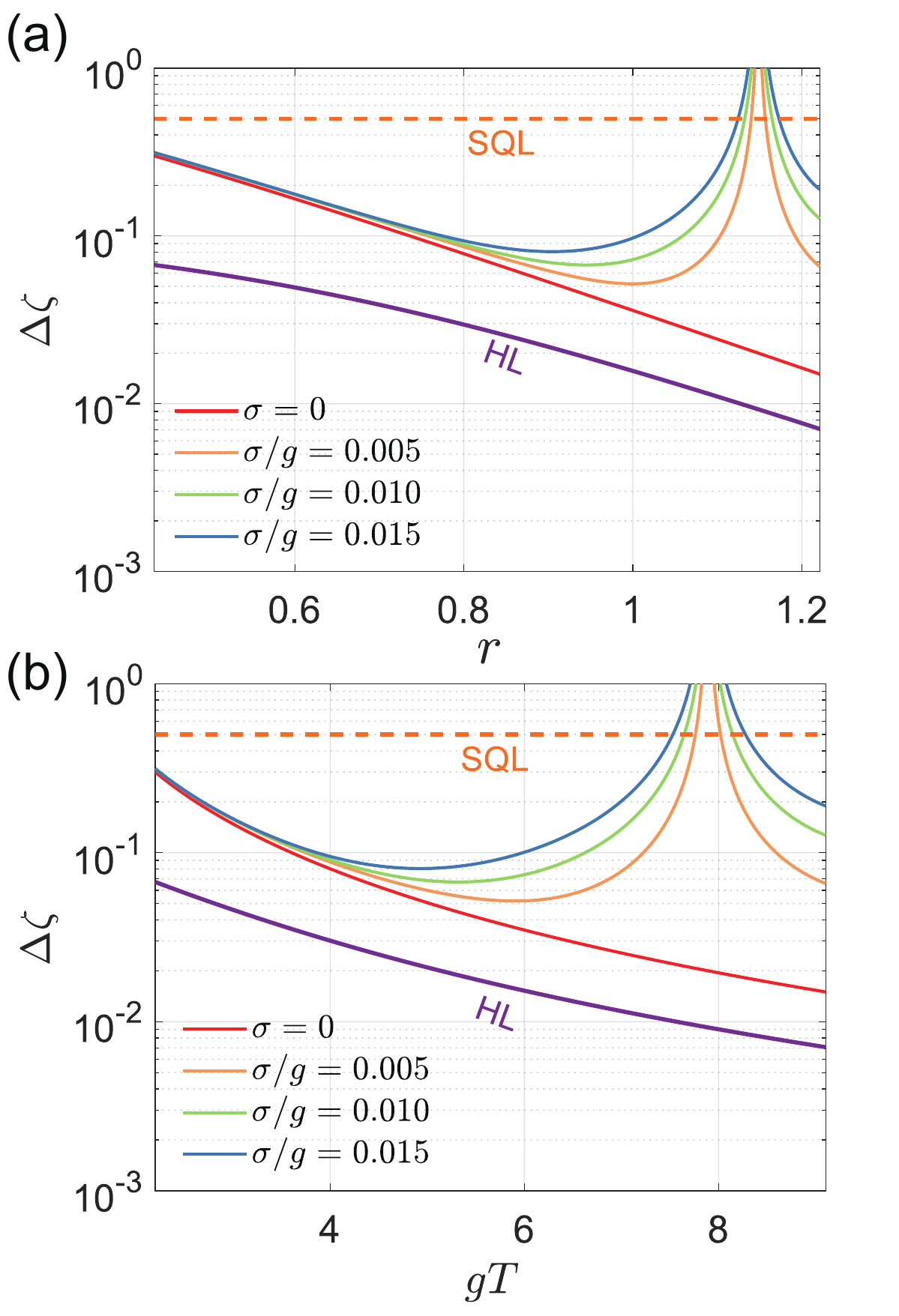}
	\caption{Displacement sensitivity $\Delta\zeta$ as a function of $r$ in (a) and $gT$ in (b) in the presence of technical noise. The red solid lines correspond to the ideal case in Eq.~\eqref{sens_nodiss_zeta}. The noise parameters are chosen as $\sigma/g=0.005,0.010,0.015$, corresponding to $\sigma/2\pi=20,40,60~{\rm Hz}$ for $g/2\pi=4~{\rm kHz}$, $\Delta_c/g=4$, $\eta=100~{\rm Hz}$ and $\Gamma=500~{\rm s}^{-1}$. The squeezing degree $r$ and the interrogation time $T$ are varied by changing $\lambda=\Omega/\Delta_c$ over $0.7\leq\lambda\leq0.985$. The orange dashed line denotes the SQL, and the purple solid line denotes the HL in Eq.~\eqref{HL_zeta}.
	}
	\label{sensing_diss}
\end{figure}
To assess the feasibility of our proposal, we now turn to the effects of technical noise on the achievable sensitivity. In this work, we focus on a trapped-ion implementation, where the dominant sources of noise  include trap-frequency fluctuations~\cite{PhysRevA.102.053106,RevModPhys.87.1419,PhysRevA.79.052104,Hu2019}, spin decoherence, and damping of the motional mode~\cite{PhysRevLett.132.220801,PhysRevA.107.032425,doi:10.1126/science.abi5226,Burd2021}. We first consider the frequency fluctuations of the motional mode. The corresponding Hamiltonian is
\begin{align}\label{freq_fluc_Hamil}
	\hat{H}_{\delta}=\;&\delta\hat{a}^\dagger\hat{a},
\end{align}
where $\delta$ is random and can be modeled as a zero-mean Gaussian variable with a variance $\sigma^2$, i.e., $\delta\sim\mathcal{N}(0,\sigma^2)$. Rewriting this Hamiltonian in terms of the Bogoliubov mode $\hat{\beta}$, we obtain~\cite{Burd2021}
\begin{align}\label{freq_fluc_Hamil_Bgl}
	\hat{H}_{\delta}=\delta\cosh(2r)\hat{\beta}^\dagger\hat{\beta}+\frac{\delta\sinh(2r)}{2}(\hat{\beta}^{\dagger2}+\hat{\beta}^2).
\end{align}
This shows that frequency fluctuations not only change the resonance frequency of the mode $\hat{\beta}$, but also generate a random squeezing term for $\hat{\beta}$. The effects of spin decoherence and motional damping on the system dynamics can be described by the Lindblad superoperator $\mathcal{L}[\hat{o}]\hat{\rho}=2\hat{o}\hat{\rho}\hat{o}^\dagger-\hat{o}^\dagger\hat{o}\rho-\hat{\rho}\hat{o}^\dagger\hat{o}$, and correspondingly, the system dynamics is governed by the master equation
\begin{align}\label{master equation spin decoherence}
	\frac{\partial\hat{\rho}}{\partial t}=-i[\hat{H}_T,\hat{\rho}]+\frac{\Gamma}{4}\mathcal{L}[\hat{\sigma}_z]\hat{\rho}+\frac{\kappa}{2}\mathcal{L}[\hat{a}]\hat{\rho}.
\end{align}
Here, $\hat{\rho}$ is the density operator of the spin-motion system, $\hat{H}_T=\hat{H}+\hat{H}_\delta$ is the total Hamiltonian of the system, and $\Gamma$ and $\kappa$ are the rates of spin decoherence and motional damping, respectively. 

Motional damping decreases the phonon population induced by the weak field $\eta$, which in turn suppresses the signal $\langle\hat{\sigma}_y\rangle_f$. In realistic experiments, however, the phonon lifetime is typically much longer than the measurement time. The effects of motional damping can therefore be safely neglected (see Appendix~\ref{Derivation of the time-dependent operator}), so that we can only consider frequency fluctuations and spin decoherence~\cite{doi:10.1126/science.abi5226,PhysRevA.107.032425}. Moreover, spin decoherence degrades the phase estimation of the spin, and as a result suppresses the signal exponentially with $\Gamma$~\cite{PhysRevLett.118.263602,PhysRevA.89.033408,PhysRevA.77.052305}. 

Specifically, for the input state $\vert+\rangle\hat{S}(re^{i\theta})\vert0\rangle$, the signal at $t=T$ is (see Appendix~\ref{Derivation of the time-dependent operator})
\begin{align}\label{final sigmay with noise}
	\langle\hat{\sigma}_y(T)\rangle\approx\exp[-\Gamma T-|A(T)|^2/2]\sin[2\Phi(T)].
\end{align}
The corresponding parameters in Eq.~\eqref{final sigmay with noise} are given by
\begin{align}\label{parameter}
	\Phi(T)=&\frac{2e^{2r}g\eta}{\xi} [-T+\frac{\sin(\upsilon T)}{\upsilon}],\\
	A(T)=&\frac{2e^r g}{\xi}\big[\cos(\upsilon T)-1+\frac{i\xi\sin(\upsilon T)}{\upsilon}\big],
\end{align}
with
\begin{align}
	\xi&=e^{2r}\delta+\omega_{\text{sq}},\label{para for parameter_a}\\
	\upsilon&=\sqrt{\delta^2+\omega_{\text{sq}}^2+2\delta\omega_{\text{sq}}\cosh(2r)}.\label{para for parameter_b}
\end{align}
Clearly, the random detuning $\delta$ modifies the accumulated phase $\Phi(T)$ and suppresses the signal $\langle\hat{\sigma}_y\rangle_f$ exponentially with a factor $|A(T)|^2/2$. According to Eq.~\eqref{sens_nodiss_zeta}, the corresponding sensitivity in our protocol is given by
\begin{align}\label{sens diss zeta}
	\Delta\zeta\approx\frac{e^{\Gamma T+|A(T)|^2/2}\sqrt{1-e^{-2\Gamma T-|A(T)|^2}\sin^2(2\Phi(T))}}{|2\cos(2\Phi(T))||\partial\Phi/\partial\zeta|},
\end{align}
for a fixed realization of the random detuning $\delta$. If $\delta=\Gamma=0$, the sensitivity $\Delta\zeta$ is reduced to Eq.~\eqref{sens_nodiss_zeta}. Throughout the numerical simulations including technical noise, we use $\sigma/(2\pi)=20$, $40$, and $60~\mathrm{Hz}$ and $\Gamma=500~\mathrm{s}^{-1}$, according to the experimental values reported in Ref.~\cite{doi:10.1126/science.abi5226}. For $g/(2\pi)=4~\mathrm{kHz}$, these values correspond to $\sigma/g=0.005$, $0.010$, and $0.015$ and $\Gamma/g=1/(16\pi)$, respectively.

\begin{figure}[t]
	\centering
	\includegraphics[width=9.0cm]{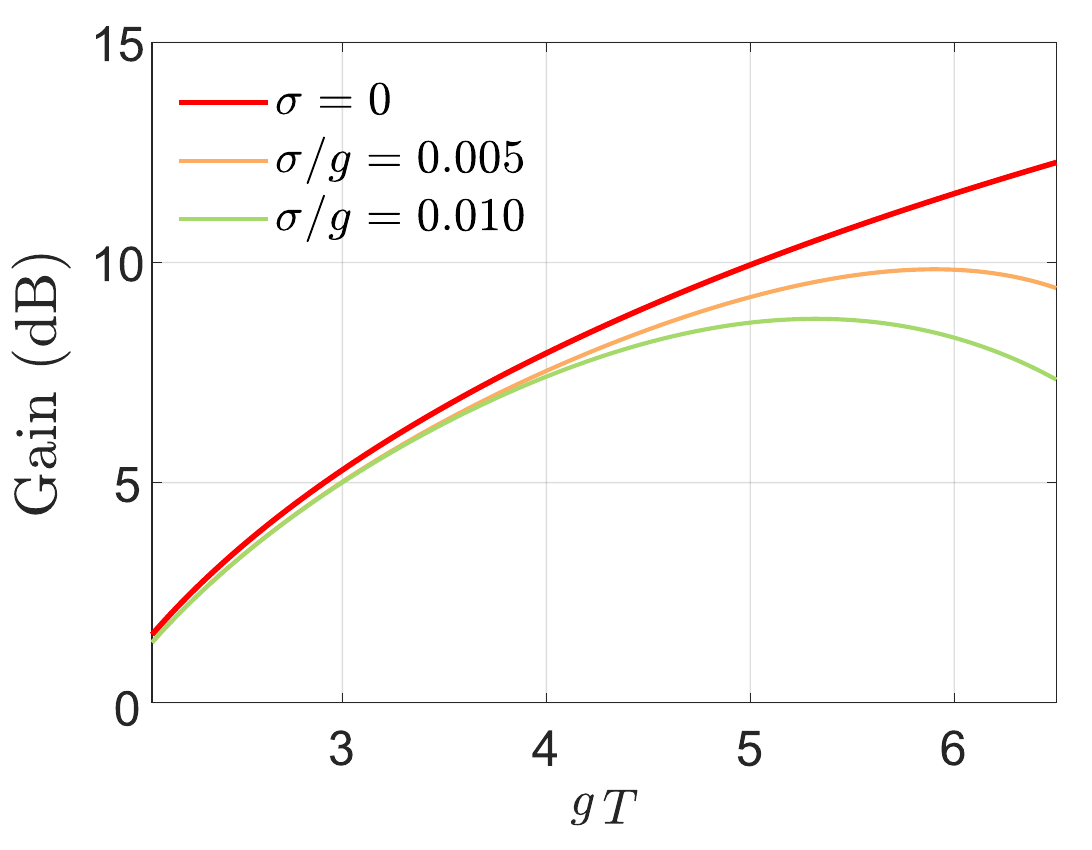}
	\caption{Metrological gain $G_{\mathrm{dB}}
		=-10\log_{10}[\Delta\zeta/(\Delta\zeta)_{\mathrm{SQL}}]$ for different frequency fluctuations $\sigma$. The other numerical parameters are the same as in Fig.~\ref{sensing_diss}.}
	\label{gain}
\end{figure}

The results plotted in Fig.~\ref{sensing_diss} are obtained by averaging the signal over the Gaussian distribution of $\delta$. The sensitivity $\Delta\zeta$ in Eq.~\eqref{sens diss zeta} shows a sharp increase near $2\Phi(T)\simeq(k+1/2)\pi$, due to the factor $\cos[2\Phi(T)]$. In the presence of technical noise, the numerator in Eq.~\eqref{sens diss zeta} remains finite, because $e^{-2\Gamma T-|A(T)|^2}<1$. At the same time, the denominator approaches zero, resulting in a sharp increase in $\Delta\zeta$. Physically, in the regime of $2\Phi(T)\simeq(k+1/2)\pi$, the final spin state is close to an eigenstate of the observable $\hat{\sigma}_y$. Note that in this regime, the signal $\langle\hat{\sigma}_y\rangle_f$ becomes insensitive to a small variation of the parameter $\zeta$, because the slope of the readout signal with respect to $\zeta$ is strongly reduced. In the ideal case with $\delta=\Gamma=0$, however,
there is no sharp degradation of the sensitivity, since the reduction of the slope is exactly compensated by the
simultaneous vanishing of the projection noise of $\hat{\sigma}_y$.

The results indicate that, even with realistic technical noise mentioned above, the sensitivity remains close to that in the absence of such noise in the small $gT$ regime. As a result, the SQL can still be surpassed for a short interrogation time $T$. Furthermore, Fig.~\ref{gain} shows the corresponding metrological gain achieved by our single-step protocol, and highlights its practical advantage in the presence of noise, with a sensitivity gain of nearly $8.7~\text{dB}$.

\section{discussions and conclusions}
\label{Conclusion}

\begin{figure}[t]
	\centering
	\includegraphics[width=9.0cm]{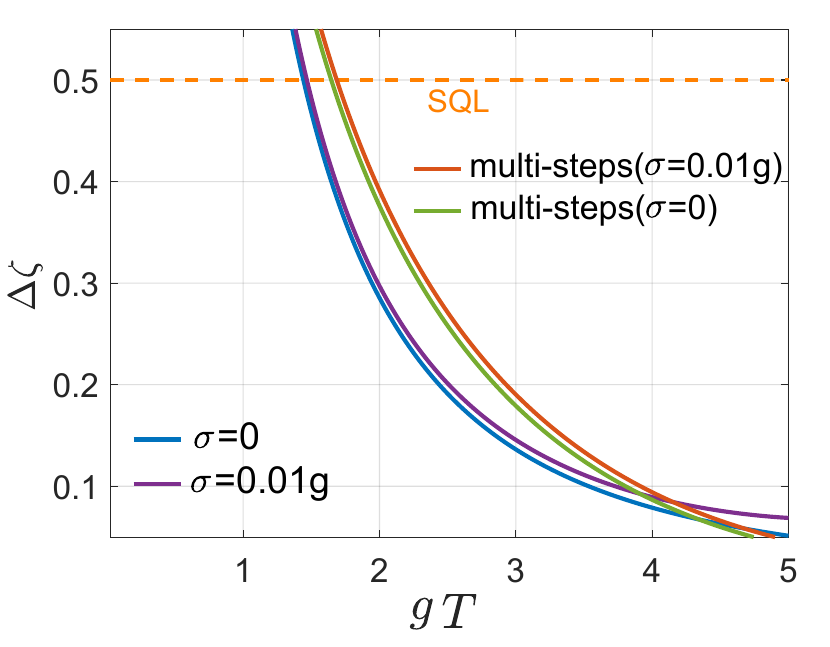}
	\caption{Comparison between our protocol and the protocol proposed in
		Ref.~\cite{PhysRevLett.132.220801}. The blue and green lines denote
		the corresponding ideal cases without dissipation. The violet and
		red lines include technical noise with
		$\Gamma/g=1/(16\pi)$ and $\sigma/g=0.010$ for both protocols,
		consistent with the noise parameters used in Fig.~\ref{sensing_diss}.
		We set $\eta=100~{\rm Hz}$ and $g/2\pi=4~{\rm kHz}$.}
	\label{compare}
\end{figure}

Finally, we discuss the metrological advantage of our protocol over previous protocols in the presence of technical noise~\cite{PhysRevLett.132.220801,PhysRevA.107.032425}. Figure~\ref{compare} compares the measurement sensitivity of our protocol with that reported in Ref.~\cite{PhysRevLett.132.220801}, which also exploits squeezing and entanglement simultaneously. For a direct comparison, we fix the DPA strength $\Omega$ and the coupling strength $g$, so that different values of $gT$ correspond to different detunings $\Delta_c$ in our protocol. As a representative example, we consider a single $^9\text{Be}^+$ ion confined in a Penning trap and set $\Omega/(2\pi)=15 \text{kHz}$ and $g/(2\pi)=4 \text{kHz}$~\cite{doi:10.1126/science.abi5226}. 

The results show that, in the short interrogation time regime, our protocol exhibits a superior performance. With a suitable choice of parameters, it can achieve a sensitivity enhancement of nearly $25\%$ even in the presence of noise. This indicates that our protocol can achieve a higher precision in the regime $gT<4$, corresponding to a shorter interrogation time, particularly with fewer operational steps. This short-time advantage arises from the fact that the squeezing strength remains constant throughout the evolution. For longer interrogation time, however, the scheme in Ref.~\cite{PhysRevLett.132.220801} eventually becomes favorable because it can reach a larger squeezing degree.

In summary, we have introduced an efficient and streamlined protocol that enhances measurement sensitivity while requiring a shorter interrogation time. In contrast to previous proposals~\cite{PhysRevLett.132.220801,PhysRevA.107.032425}, our scheme relies on only one detuned motional squeezing step and does not require any anti-squeezing process. This substantially simplifies the experimental procedure, reduces operational complexity, and helps mitigate additional errors in practical implementations. By combining squeezing and entanglement, our scheme simultaneously amplifies both the signal and the coupling strength during the interrogation. Moreover, by employing a fixed detuning term, the interrogation time can be set to an optimal value, such that the spin and motional degrees of freedom are disentangled completely at the end of the interrogation. Although the present analysis is restricted to trapped-ion systems~\cite{PhysRevLett.122.030501,PhysRevA.42.2977}, the proposed protocol is not limited to this platform and can, in principle, be extended to other systems, such as circuit QED~\cite{PRXQuantum.6.010304} and NV centers~\cite{PhysRevApplied.20.034063}. These features make it a promising scheme for fast and high-precision weak-signal sensing, particularly in experimental scenarios where the available interrogation time is limited by decoherence and technical noise.

\begin{acknowledgments}
This work was supported by the National Natural Science Foundation of China (NSFC) via Grants No.~0401260012 and No.~12404419.
\end{acknowledgments}
\appendix %%MDPI internal note: new layout%%

\setcounter{equation}{0} \setcounter{figure}{0}
\setcounter{table}{0} \makeatletter
\renewcommand{\thefigure}{A\arabic{figure}}

\section{Derivation of the state in Eq.~\eqref{state_with_t}}
\label{Derivation of the state}

In this section, we provide the derivation of the time-dependent state in Eq.~\eqref{state_with_t}. We also relate the Bogoliubov-mode displacement to the bare-mode displacement used in the main text. 

\subsection{Detailed derivation of Eq.~\eqref{state_with_t}}
Starting from Eq.~\eqref{Bogoliubov Hamiltonian}, we have
\begin{equation}
	\hat H_{\rm sq}
	=
	\omega_{\rm sq}\hat\beta^\dagger\hat\beta
	+
	e^r(g\hat\sigma_z+\eta)(\hat\beta+\hat\beta^\dagger).
\end{equation}
Using
$\hat D_\beta^\dagger(\hat f)\hat\beta\hat D_\beta(\hat f)
=\hat\beta+\hat f$, with
\begin{equation}
	\hat f
	=
	-\frac{e^r(g\hat\sigma_z+\eta)}{\omega_{\rm sq}},
\end{equation}
we obtain
\begin{align}
	\hat H'_{\mathrm{sq}}
	&=
	\hat D_\beta^\dagger(\hat f)
	\hat H_{\mathrm{sq}}
	\hat D_\beta(\hat f)
	\nonumber\\
	&=
	\omega_{\mathrm{sq}}
	\hat\beta^\dagger\hat\beta
	-
	\frac{2g\eta e^{2r}}
	{\omega_{\mathrm{sq}}}
	\hat\sigma_z
	-
	\frac{e^{2r}(g^2+\eta^2)}
	{\omega_{\mathrm{sq}}}.
\end{align}
The last term contributes only a global phase and is therefore omitted in the below derivations. The inverse transformation gives
\begin{equation}
	\hat H_{\rm sq}
	=
	\hat D_\beta(\hat f)
	\hat H_{\rm sq}'
	\hat D_\beta^\dagger(\hat f).
\end{equation}
Consequently, the time-evolution operator is
\begin{equation}
	\hat U_{\rm sq}(t)
	=
	e^{\frac{2i\eta ge^{2r}t\hat\sigma_z}{\omega_{\rm sq}}}
	\hat D_\beta(\hat f)
	e^{-i\omega_{\rm sq}\hat\beta^\dagger\hat\beta t}
	\hat D_\beta^\dagger(\hat f).
\end{equation}
Using the Baker-Campbell-Hausdorff formula, we obtain
\begin{align}
	&\hat D_\beta(\hat f)
	e^{-i\omega_{\rm sq}\hat\beta^\dagger\hat\beta t}
	\hat D_\beta^\dagger(\hat f)
	\nonumber\\
	&\quad=
	e^{-i\hat f^2\sin(\omega_{\rm sq}t)}
	\hat D_\beta\left[
	\hat f(1-e^{-i\omega_{\rm sq}t})
	\right]
	e^{-i\omega_{\rm sq}\hat\beta^\dagger\hat\beta t}.
\end{align}
After combining the spin-dependent phases, the evolution operator becomes
\begin{equation}\label{time_evo_oper}
	\hat U_{\rm sq}(t)
	=
	e^{i\phi(t)\hat\sigma_z}
	\hat D_\beta[\hat{\alpha}(t)]
	e^{-i\omega_{\rm sq}\hat\beta^\dagger\hat\beta t},
\end{equation}
with
\begin{align}
	\hat{\alpha}(t)
	&=
	-\frac{e^r(\eta+g\hat\sigma_z)}
	{\omega_{\rm sq}}
	\left(1-e^{-i\omega_{\rm sq}t}\right),\\
	\phi(t)
	&=
	\frac{2\eta ge^{2r}}
	{\omega_{\rm sq}^2}
	\left[
	\omega_{\rm sq}t-\sin(\omega_{\rm sq}t)
	\right].
\end{align}
Then, applying this evolution operator to the initial state $|+\rangle|\varphi_0\rangle$ yields the time-dependent state in Eq.~\eqref{state_with_t}. At $t=T$, we have $\hat\alpha(T)=0$ and $e^{-i\omega_{\rm sq}\hat\beta^\dagger\hat\beta T}=\hat I$.
Therefore, the motional degree of freedom returns to its initial state. At the same time, the accumulated spin phase is enhanced by a factor of $e^{2r}$, which is
independent of the initial motional state.

To evaluate the mean phonon number independently of the unknown signal, we set $\eta=0$. For the squeezed-vacuum initial motional state $\hat{S}(re^{i\theta})\vert0\rangle$, the corresponding spin-motion state at $t=T/2$ is
\begin{equation}
	|\Psi\rangle
	=
	\frac{1}{\sqrt{2}}
	\left[
	|\uparrow\rangle\hat D_a(-x)\hat S(re^{i\theta})|0\rangle
	+
	|\downarrow\rangle\hat D_a(x)\hat S(re^{i\theta})|0\rangle
	\right],
\end{equation}
where $x=gTe^{2r}/\pi$ is the bare-mode displacement amplitude. This is the intermediate state in Eq.~\eqref{Psi} used to calculate $\bar n_{\max}$. 

\subsection{Case of an arbitrary signal-field phase}
\label{Alternative phase convention}
For an arbitrary signal phase $\phi_{E}$, the signal Hamiltonian in the rotating frame can be written as~\cite{Wolf2019}
\begin{equation}
	\hat H_E
	=
	\eta\left(
	e^{-i\phi_E}\hat a
	+
	e^{i\phi_E}\hat a^\dagger
	\right).
\end{equation}
In this case, we set $\phi_D=\phi_E$, so that the ODF Hamiltonian becomes
\begin{equation}
	\hat H_{\rm ODF}
	=
	g\left(
	e^{-i\phi_E}\hat a
	+
	e^{i\phi_E}\hat a^\dagger
	\right)\hat\sigma_z.
\end{equation}
We further choose the DPA phase as $\theta=\pi+2\phi_E$. The corresponding Bogoliubov transformation gives
\begin{equation}
	e^{-i\phi_E}\hat a
	+
	e^{i\phi_E}\hat a^\dagger
	=
	e^r\left(
	e^{-i\phi_E}\hat\beta
	+
	e^{i\phi_E}\hat\beta^\dagger
	\right),
\end{equation}
and the Hamiltonian in Eq.~\eqref{Bogoliubov Hamiltonian} becomes
\begin{equation}
	\hat H_{\rm sq}
	=
	\omega_{\rm sq}\hat\beta^\dagger\hat\beta
	+
	e^r(g\hat\sigma_z+\eta)
	\left(
	e^{-i\phi_E}\hat\beta
	+
	e^{i\phi_E}\hat\beta^\dagger
	\right).
\end{equation}
Then, introducing
\begin{equation}
	\hat f_{\phi_E}
	=
	-\frac{e^r(g\hat\sigma_z+\eta)e^{i\phi_E}}
	{\omega_{\rm sq}},
\end{equation}
the time-evolution operator $\hat{U}_{\rm sq}(t)$, corresponding to the arbitrary-signal-phase extension of Eq.~\eqref{time_evo_oper}, is given by
\begin{equation}
	\hat U_{\rm sq}(t)
	=
	e^{i\phi(t)\hat\sigma_z}
	\hat D_\beta[\alpha_{\phi_E}(t)]
	e^{-i\omega_{\rm sq}\hat\beta^\dagger\hat\beta t},
\end{equation}
where
\begin{align}
	\hat \alpha_{\phi_E}(t)
	&=
	-\frac{e^r(g\hat\sigma_z+\eta)e^{i\phi_E}}
	{\omega_{\rm sq}}
	\left(1-e^{-i\omega_{\rm sq}t}\right),\\
	\phi(t)
	&=
	\frac{2g\eta e^{2r}}{\omega_{\rm sq}^2}
	\left[
	\omega_{\rm sq}t-\sin(\omega_{\rm sq}t)
	\right].
\end{align}
At $T=2\pi/\omega_{\rm sq}$, we have
$\hat \alpha_{\phi_E}(T)=0$, and the final state is the same as in Eq.~\eqref{reduced final state}. This means that the exponential enhancement, as given in Eq.~\eqref{e2r}, of the accumulated phase persists for an arbitrary signal-field phase $\phi_{E}$.

\section{Quantum Fisher information and displacement Heisenberg limit}
\label{Derivation of the quantum Fisher information}

In this section, we derive the time-dependent QFI in Eq.~\eqref{time-dependent QFI} and establish the displacement Heisenberg limit used in the main text.

\subsection{Time-dependent QFI in Eq.~\eqref{time-dependent QFI}}
For the initial motional state $|\varphi_0\rangle=\hat{S}(re^{i\theta})\vert0\rangle=\vert0\rangle_\beta$, the time-dependent state and its derivative are
\begin{align}
	|&\psi(t)\rangle
	=
	e^{i\phi(t)\hat\sigma_z}
	\hat D_\beta[\hat \alpha(t)]
	|+\rangle|0\rangle_\beta,\\
	|\partial_\eta&\psi(t)\rangle
	={}
	e^{i\phi(t)\hat\sigma_z}
	\hat D_\beta[\hat{\alpha}(t)]
	\nonumber\\
	&\times
	\Big[
	i(\partial_\eta\phi(t))\hat\sigma_z
	+(\partial_\eta\hat{\alpha}(t))\hat\beta^\dagger
	-(\partial_\eta\hat{\alpha}^\dagger(t))\hat\beta
	\Big]
	|+\rangle|0\rangle_\beta,
\end{align}
where $\hat{\alpha}(t)$ and $\phi(t)$ are given by Eqs.~\eqref{alpha(t)} and \eqref{phi(t)}, respectively. The derivatives of $\hat{\alpha}(t)$ and $\phi(t)$ with respect to $\eta$ are
\begin{align}
	\partial_\eta\hat{\alpha}(t)
	&=
	-\frac{e^r}{\omega_{\rm sq}}
	\left(1-e^{-i\omega_{\rm sq}t}\right),\\
	\partial_\eta\phi(t)
	&=
	\frac{2ge^{2r}}{\omega_{\rm sq}^2}
	\left[
	\omega_{\rm sq}t-\sin(\omega_{\rm sq}t)
	\right].
\end{align}
Using these results, we obtain
\begin{align}
	\langle\psi(t)|
	\partial_\eta\psi(t)\rangle
	&=0,\\
	\langle\partial_\eta\psi(t)|
	\partial_\eta\psi(t)\rangle
	&=
	|\partial_\eta\hat{\alpha}(t)|^2
	+
	[\partial_\eta\phi(t)]^2.
\end{align}
Substituting these results into Eq.~\eqref{QFI} yields
\begin{align}
	\mathcal{F}^\eta_Q(t)
	={}&
	\frac{16e^{2r}}{\omega_{\rm sq}^2}
	\sin^2\left(\frac{\omega_{\rm sq}t}{2}\right)
	\nonumber\\
	&+
	\frac{16g^2e^{4r}}{\omega_{\rm sq}^4}
	\left[
	\omega_{\rm sq}t-\sin(\omega_{\rm sq}t)
	\right]^2,
\end{align}
which is the QFI in Eq.~\eqref{time-dependent QFI}.

\subsection{Heisenberg limit}
Consider an arbitrary pure input state $\vert\psi\rangle$ with the displacement encoding
\begin{equation}
	|\psi_\zeta\rangle
	=
	e^{-i\zeta\hat H_{\rm d}}|\psi\rangle,
	\qquad
	\hat H_{\rm d}=\hat a+\hat a^\dagger.
\end{equation}
The corresponding QFI satisfies
\begin{equation}
	\mathcal F_Q^{\rm disp}
	=
	4(\Delta\hat H_{\rm d})^2
	\leq
	4\langle\psi|\hat H_{\rm d}^2|\psi\rangle.
\end{equation}
Defining
\begin{equation}
	\bar n
	=
	\langle\psi|\hat a^\dagger\hat a|\psi\rangle,
\end{equation}
we have
\begin{equation}
	\langle\psi|\hat{H}_{\rm d}^2|\psi\rangle
	=
	2\bar n+1
	+2\operatorname{Re}
	\langle\psi|\hat a^2|\psi\rangle.
\end{equation}
Applying the Cauchy-Schwarz inequality to $\hat a^\dagger|\psi\rangle$ and $\hat a|\psi\rangle$ gives
\begin{align}
	|\langle\psi|\hat a^2|\psi\rangle|^2
	&\leq
	\langle\psi|\hat a\hat a^\dagger|\psi\rangle
	\langle\psi|\hat a^\dagger\hat a|\psi\rangle
	\nonumber\\
	&=
	\bar n(\bar n+1).
\end{align}
Using the above inequality, the QFI satisfies
\begin{align}
	\mathcal F_Q^{\rm disp}
	&\leq4\left[2\bar n+1+2\sqrt{\bar n(\bar n+1)}\right]
	\nonumber\\
	&\leq16\bar n+8,
\end{align}
where the relation
$\sqrt{\bar n(\bar n+1)}\leq\bar n+1/2$ has been used. Taking $\bar n_{\max}$ as the maximum mean phonon number during the sensing process, we obtain
\begin{equation}
	\mathcal F_Q^{\rm disp}
	\leq
	16\bar n_{\max}+8.
\end{equation}
Using Eqs.~\eqref{time-dependent QFI} and~\eqref{phonon number}, one can further verify that the QFI of our protocol satisfies
\begin{equation}
	\mathcal{F}^\zeta_Q(t)
	\leq
	16\bar n_{\max}+8,
	\qquad 0\leq t\leq T.
\end{equation}
The corresponding sensitivity bound is
\begin{equation}
	\Delta\zeta\geq\frac{1}{4\sqrt{\bar n_{\max}+1/2}}.
\end{equation}
We refer to this lower bound as the Heisenberg limit for displacement sensing and use it as the Heisenberg-limit benchmark in the main text~\cite{mhvv-pr5d}.

\section{Derivation of the expectation value in Eq.~\eqref{final sigmay with noise}}
\label{Derivation of the time-dependent operator}

\begin{figure}[t]
	\centering
	\includegraphics[width=9.0cm]{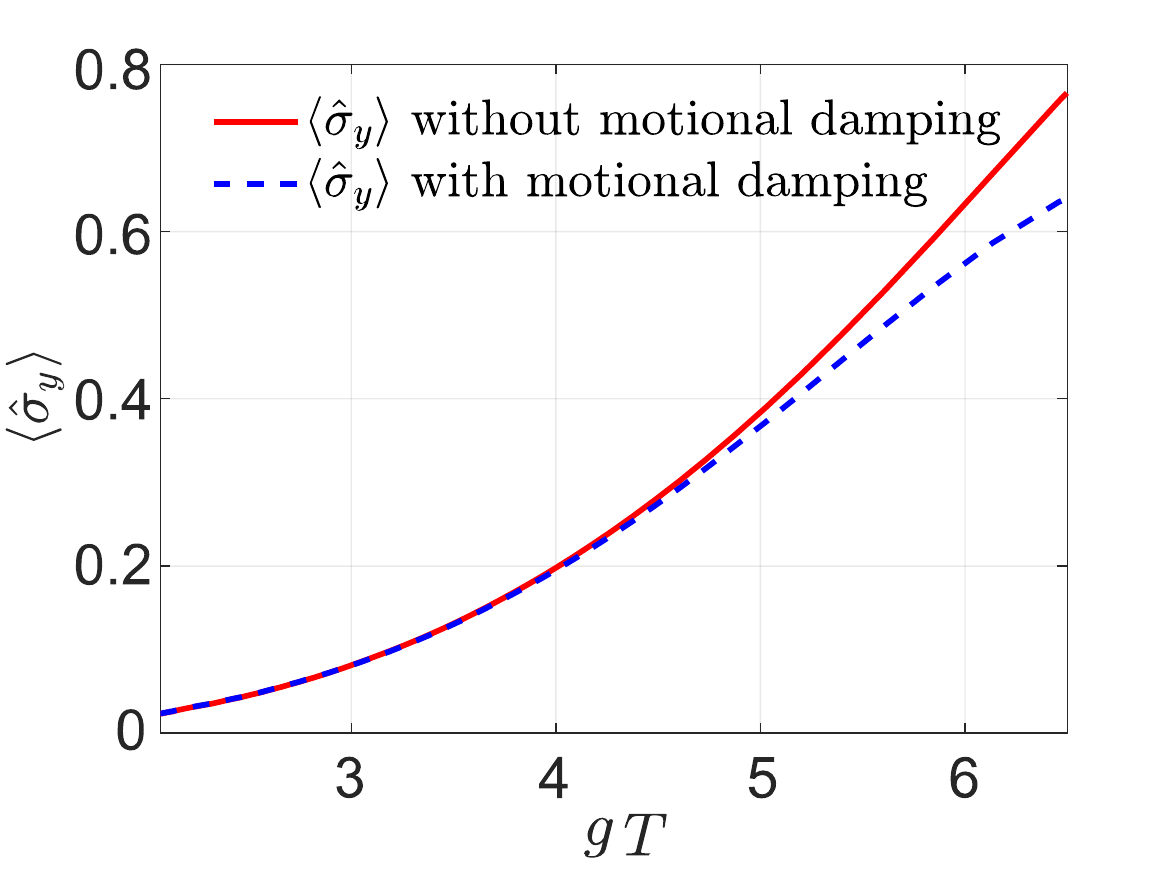}
	\caption{Expectation value $\langle\hat{\sigma}_y\rangle$ with (dashed line) and without (solid line) motional damping. The decay rate $\kappa=(1/300\text{ms})$, and the other parameters are the same as in Fig.~\ref{compare}.}
	\label{sigmay_diss}
\end{figure}

In this section, we provide a detailed derivation of the expectation value $\langle\hat{\sigma}_y(t)\rangle$. 

As shown in Fig.~\ref{sigmay_diss}, in the short time regime, the motional damping scarcely affects the signal, even though squeezing enhances the dissipation. The effect of motional damping is thus negligible in our scheme. Then, using the Heisenberg equations of motion, we obtain the time evolution of $\hat{\beta}(t)$ and $\hat{\sigma}_+(t)$ as follows~\cite{doi:10.1126/science.abi5226,PhysRevA.107.032425}:
\begin{align}\label{Heisenberg_equ}
	i\dot{\hat{\beta}}(t)&=\chi\hat{\beta}(t)+(g\hat{\sigma}_z+\eta)e^r+\delta\sinh(2r)\hat{\beta}^\dagger(t),\nonumber\\
	-i\dot{\hat{\beta}}^\dagger(t)&=\chi\hat{\beta}^\dagger(t)+(g\hat{\sigma}_z+\eta)e^r+\delta\sinh(2r)\hat{\beta}(t),\nonumber\\
	i\dot{\hat{\sigma}}_+(t)&=-2ge^r(\hat{\beta}(t)+\hat{\beta}^\dagger(t))\hat{\sigma}_+(t),
\end{align}
here we set $\chi=\omega_{\text{sq}}+\delta\cosh(2r)$. Note that spin decoherence can be taken into account independently of the frequency fluctuations and contributes only an overall decay factor $e^{-\Gamma T}$ to the signal. From the above equations, the operator $\hat{\sigma}_+(t)$ is found to be
\begin{align}\label{freq_fluc_sig+}
		\hat{\sigma}_+(t)=\;&\exp[\frac{2ig\Phi(t)}{\eta}(\hat{\sigma}_z-1)]\exp[2i\Phi(t)]\hat{D}_\beta[A(t)]\hat{\sigma}_+(0),
\end{align}
where $\hat{D}_\beta(\alpha)=e^{\alpha\hat{\beta}^\dagger-\alpha^*\hat{\beta}}$ is the displacement operator in $\hat{\beta}$ mode, the corresponding parameters are given by
\begin{align}\label{freq_fluc_sig+_para}
		\Phi(t)=&\frac{2e^{2r}g\eta}{\xi} [-t+\frac{\sin(\upsilon t)}{\upsilon}],\\
		A(t)=&\frac{2e^r g}{\xi}\big[\cos(\upsilon t)-1+\frac{i\xi\sin(\upsilon t)}{\upsilon}\big],
\end{align}
where $\xi$ and $\upsilon$ are given in Eqs.~\eqref{para for parameter_a} and \eqref{para for parameter_b}. The expectation value $\langle\hat{\sigma}_y(t)\rangle$ in Eq.~\eqref{final sigmay with noise} is then obtained according to 
\begin{align}\label{for sy}
	\langle\hat{\sigma}_y(t)\rangle=-i(\langle\hat{\sigma}_+(t)\rangle-\langle\hat{\sigma}_-(t)\rangle).
\end{align}

\end{document}